\documentclass[eqsecnum,superscriptaddress,nofootinbib,11pt]{revtex4}
\usepackage{amsmath}
   \usepackage{bm}
\usepackage{amsthm,amscd}
\usepackage{mathrsfs}
\usepackage{verbatim}
\usepackage{amsfonts,amsmath,latexsym,amssymb,txfonts,bm}
\usepackage[american]{babel}
\usepackage{dsfont}
\usepackage[dvips]{graphicx}
\usepackage{latexsym}
\usepackage{epsfig}
\newcommand{\diff}{\mathrm{d}}

\def\beq{\begin{eqnarray}}
\def\eeq{\end{eqnarray}}

\begin{document}
\title{Functional determinants and Casimir energy in higher dimensional spherically symmetric background potentials}

\author{Guglielmo Fucci\footnote{Electronic address: fuccig@ecu.edu}}
\affiliation{Department of Mathematics, East Carolina University, Greenville, NC 27858 USA}

\author{Klaus Kirsten\footnote{Electronic address: Klaus\textunderscore Kirsten@Baylor.edu}}
\affiliation{GCAP-CASPER, Department of Mathematics, Baylor University, Waco, TX 76798 USA}

\date{\today}
\vspace{2cm}
\begin{abstract}

In this paper we analyze the spectral zeta function associated with a Laplace operator acting on scalar functions on an $N$-dimensional Euclidean space
in the presence of a spherically symmetric background potential. The obtained analytic continuation of the
spectral zeta function is then used to derive very simple results for the functional determinant of the operator and the Casimir energy
of the scalar field.

\end{abstract}
\maketitle

\section{Introduction}

The detailed analysis of the influence that non-dynamical external fields have on quantized fields is a very important and interesting
component of quantum field theory. Non-dynamical external fields arise naturally in models describing the dynamics of quantum fields
in the presence of background potentials resulting from classical solutions such as monopoles \cite{hoof74-79-276,poly74-20-194},
sphalerons \cite{klin84-30-2212} and electroweak Skyrmions \cite{gips81-183-524,gips84-231-365,ambj85-256-434,eila86-56-1331,frie77-15-1694,frie77-16-1096,skyr61-260-127,skyr62-31-556,adki83-228-552}.
The spectral zeta function has been proven to be an extremely valuable tool in many areas of mathematics
and physics \cite{elizalde94,gilkey95,kirs02b} especially for the study of the one-loop effective action in quantum field theory \cite{bytsenko03} and the vacuum energy \cite{bord09b}. For this reason, spectral zeta function regularization techniques have been widely utilized to approach
the types of problems involving the interaction between quantum fields and non-dynamical external fields described above.
The dynamics of a quantum field under the presence of a non-dynamical spherically symmetric external field is described by a
Laplace operator endowed with a spherically symmetric potential.
The appropriate formalism needed to analyze the spectral zeta function associated with this operator
was developed in \cite{bordag96,dunne09}. Within this framework the spectral zeta function
is represented in terms of a complex integral. The path of integration is then deformed along the imaginary axis so as to
obtain an integral over imaginary frequencies. The analytic continuation of the spectral zeta function to a meromorphic function in the complex plane
is obtained by adding and subtracting the asymptotic expansion of the logarithm of the Jost functions \cite{bordag96}.
The asymptotic terms and the logarithm of the Jost function with the subtracted asymptotic terms are first integrated over the imaginary frequency and
the obtained expressions are then summed over the angular momenta. The procedure just described
consists of a series of calculations that are quite involved. Moreover, the analytically continued expression for the
spectral zeta function and the information one can extract from it, such as the vacuum energy of a quantum field,
are usually given by somewhat complex formulas.

In order to overcome some of the inherent technical difficulties that are present in the approach developed in \cite{bordag96,dunne09},
a different method for the analytic continuation of the spectral zeta function for Laplace operators endowed with spherically symmetric potentials
has been proposed in \cite{bea15}; see also \cite{moss02-632-173}. In this new approach the sum over the angular momenta of the integral representation of the spectral zeta function
is performed {\em before} the integration over the imaginary frequency. Within this framework, particularly critical roles are played by
the phase shift and the asymptotic expansion of the trace of the heat kernel associated with the Laplace type operator $-\Delta+V(r)+m^{2}$ endowed with a spherically symmetric  potential $V(r)$.
The improved technique has been used to obtain surprisingly simple results for the vacuum energy of scalar fields under the
influence of spherically symmetric potentials in a two and three dimensional Euclidean space. In this paper we extend the
results obtained in \cite{bea15} for the vacuum energy of a scalar field to higher dimensions. In addition, we also provide an expression for the functional
determinant of the operator $-\Delta+V(r)+m^{2}$ acting on functions defined on $\mathbb{R}^{D}$. The additional dimensions that we consider in this work
lead to calculations that are more involved that the ones presented in \cite{bea15} due to the more complicated structure of the angular momenta
in higher dimension.

The outline of the paper is as follows. In the next section we introduce the spectral zeta function of the problem
under consideration and provide its integral representation in terms of the phase shift. In Section III the analytic continuation of the
zeta function is obtained in terms of the asymptotic expansion of the trace of the heat kernel of the Laplacian under consideration.
The subsequent sections provide explicit results for the Casimir energy of a scalar field and the functional determinant
of the operator. The last section summarizes the main results of the paper and points to a few directions for further investigation.

\section{The spectral zeta function}

In this paper we consider the following Laplace-type operator
\begin{equation}\label{0}
{\cal L}=-\Delta+V(r)+m^{2}\;,
\end{equation}
acting on scalar functions defined on the Euclidean space $\mathbb{R}^{D}$. In the above expression, $\Delta$ denotes the
familiar Laplacian in $\mathbb{R}^{D}$, the function $V(r)$ represents a spherically symmetric potential which decays sufficiently fast as $r\to\infty$, with $r$ being the radial
coordinate, and $m>0$ is the mass of the scalar field. The non-trivial part of the eigenvalue equation associated with the operator (\ref{0}) is
\begin{equation}\label{1}
{\cal P}\phi_{j}=\lambda^{2}_{j}\phi_{j}\;.
\end{equation}
For the moment, we will assume that the scalar field $\phi$ is confined within a sphere of radius $R$ and satisfies
Dirichlet boundary conditions $\phi(R)=0$. Under this assumption, equation (\ref{1}) together with the aforementioned boundary condition
leads to a discrete set of eigenvalues $\lambda_{j}^2$. In this case the index $j\in\mathbb{N}^{+}$ and the spectral
zeta function associated with the operator ${\cal L}$, including the trivial mass dependence, is given as
\begin{equation}\label{2}
\zeta(s)=\sum_{j=1}^{\infty}\left(\lambda_{j}^{2}+m^{2}\right)^{-s}\;,
\end{equation}
which is well defined for $\Re(s)>D/2$. In this work we will be mainly concerned with the evaluation of both the
Casimir energy for the scalar field $\phi$ under the influence of the potential $V(r)$ and the functional determinant of the operator ${\cal L}$.
It is well known that the spectral zeta function can be utilized to compute the Casimir energy $E_{\textrm{Cas}}$ of a quantum field through the
following formula \cite{bord09b,kirs02b}
\begin{equation}\label{3}
E_{\textrm{Cas}}=\lim_{\epsilon\to 0}\frac{\mu^{2s}}{2}\zeta\left(\epsilon-\frac{1}{2}\right)\;,
\end{equation}
where $\mu$ represents a parameter with the dimension of a mass. Moreover, within the framework of spectral zeta
function regularization, the functional determinant of ${\cal L}$ is defined according to the expression \cite{ray71,hawk77-55-133,dowk76-13-3224}
\begin{equation}\label{4}
\textrm{Det}\,{\cal L}=\exp\left\{-\zeta'(0)\right\}\;.
\end{equation}
Since the formula for the spectral zeta function in equation (\ref{2}) is only valid for $\Re(s)>D/2$, the definitions in
(\ref{3}) and (\ref{4}) require the use of the analytically continued expression of $\zeta(s)$. To this end, it can
be proved \cite{mina49,seeley,voros87} that the spectral zeta function can be analytically continued to the entire complex
plane to a meromorphic function possessing only isolated simple poles.

In this work we first rewrite the spectral zeta function in (\ref{2}) in terms of a contour integral in the complex plane
and use that integral representation as the starting point for the analytic continuation. This technique has been used successfully
several times in the literature in a wide variety of settings (see e.g. \cite{kirs02b,kirs03-308-502} for a primer on the subject).

By using spherical coordinates $(r,\boldsymbol{\theta})$, the equation (\ref{1}) can be solved by separation and the solution
can be written as a product $\phi_{l,p}=r^{-\frac{D-1}{2}}\omega_{l,p}(r)\Psi_{l}(\boldsymbol{\theta})$. Here, the angular
function $\Psi_{l}(\boldsymbol{\theta})$ satisfies the equation
\begin{equation}\label{5}
-\Delta_{S^{D-1}}\Psi_{l}(\boldsymbol{\theta})=l(l+D-2)\Psi_{l}(\boldsymbol{\theta})\;,
\end{equation}
with $l\in\mathbb{N}_{0}$ and $\Delta_{S^{D-1}}$ being the Laplacian on the $(D-1)$-dimensional sphere, and
the radial function $\omega_{l,p}(r)$ is a solution of
\begin{equation}\label{6}
\left\{\frac{\diff^{2}}{\diff r^{2}}-\frac{1}{r^{2}}\left[\left(l-1+\frac{D}{2}\right)^{2}-\frac{1}{4}\right]-V(r)+p^{2}\right\}\omega_{l,p}(r)=0\;,
\end{equation}
with $p^{2}$ denoting a positive parameter we use in lieu of the eigenvalue $\lambda_{j}^{2}$.
An explicit solution for the radial equation (\ref{6}) and the parameter $p^{2}$ (which would then coincide with the eigenvalue $\lambda_{j}^{2}$) can only be found
for very few special potentials $V(r)$. However, for the purpose of performing the analytic continuation of the spectral zeta function
this explicit knowledge is not necessary, in fact the information one can obtain from scattering theory proves to be sufficient.

As $r\to\infty$ the scattered waves become free, and the solutions of equation (\ref{6}) have to be compared with
the solutions of the free equation, obtained from (\ref{6}) by setting $V(r)=0$, namely
\begin{equation}\label{7}
y_{l,p}(r)=r^{1/2}\left[C_{1}J_{l+\frac{D-2}{2}}(pr)+C_{2}Y_{l+\frac{D-2}{2}}(pr)\right]\;,
\end{equation}
where $(C_{1},C_{2})$ are arbitrary constants and $J_{\nu}(x)$ and $Y_{\nu}(x)$ represent the Bessel functions of the first and second kind, respectively.
The only physically relevant solution, however, is the one that behaves like $r^{l+\frac{D-1}{2}}$ as $r\to 0$ which can be easily found to be the Riccati-Bessel function
\begin{equation}\label{8}
y_{l,p}(r)=\hat{j}_{l+\frac{D-3} 2}(pr)=\sqrt{\frac{\pi p r}{2}}J_{l+\frac{D-2}{2}}(pr)\;.
\end{equation}
In light of the above discussion we define the regular solution $\phi_{l,p}(r)$ of (\ref{6}) as the one that is proportional to $\omega_{l,p}(r)$ and such that
\begin{equation}\label{9}
\phi_{l,p}(r)\sim \hat{j}_{l+\frac{D-3} 2}(pr)\;,
\end{equation}
as $r\to 0$.
As $r\to\infty$, instead, the regular solution must behave as a linear combination of free outgoing and incoming waves, namely
\begin{equation}\label{10}
\phi_{l,p}(r)\sim \frac{i}{2}\left[f_{l}(p)\hat{h}^{-}_{l+\frac{D-3} 2}(pr)-f^{\ast}_{l}(p)\hat{h}^{+}_{l+\frac{D-3} 2}(pr)\right]\;,
\end{equation}
where the coefficients $f_{l}(p)$ and $f^{\ast}_{l}(p)$ represent the Jost function and its complex conjugate, respectively, and
$\hat{h}^{-}_{l}(pr)$ and $\hat{h}^{+}_{l}(pr)$ are the Riccati-Hankel functions
\begin{equation}
\hat{h}^{+}_{l+\frac{D-3} 2}(pr)=i\sqrt{\frac{\pi p r}{2}}H^{(1)}_{l+\frac{D-2}{2}}(pr)\;,\quad\textrm{and}\quad
\hat{h}^{-}_{l+\frac{D-3} 2}(pr)=-i\sqrt{\frac{\pi p r}{2}}H^{(2)}_{l+\frac{D-2}{2}}(pr)\;.
\end{equation}

Let us assume, for now, that the potential $V(r)$ has a compact support contained in the sphere of radius $R$ \cite{bordag96}.
In this case for $r=R$ the regular solution in (\ref{10}) becomes exact and by imposing Dirichlet boundary conditions, namely $\phi_{l,p}(R)=0$,
we obtain an implicit equation for the eigenvalues $\lambda_{j}^2$ of (\ref{1}). More precisely one has
\begin{equation}\label{11}
f_{l}(p)\hat{h}^{-}_{l+\frac{D-3} 2}(pR)-f^{\ast}_{l}(p)\hat{h}^{+}_{l+\frac{D-3} 2}(pR)=0\;.
\end{equation}
We would like to mention that imposing boundary conditions at $r=R$ represents only an intermediate technical step.  We are actually
mainly interested in the limit as $R\to\infty$. In this limit the results obtained later will be independent on the choice of boundary conditions
once we assume that $V(r)\sim r^{-D+1-\varepsilon}$ with $\varepsilon>0$ for $r\to\infty$ \cite{bordag96}.
We can utilize the above implicit equation to write an integral representation for the spectral zeta function \cite{bordag96}.
By subtracting the contributions arising from the free Euclidean space we obtain
\begin{equation}\label{12}
\zeta(s)=\frac{1}{2\pi i}\sum_{l=0}^{\infty}d(l)\int_{\gamma}\left(p^{2}+m^{2}\right)^{-s}\frac{\partial}{\partial p}\log\left[\frac{f_{l}(p)\hat{h}^{-}_{l+\frac{D-3} 2}(pR)-f^{\ast}_{l}(p)\hat{h}^{+}_{l+\frac{D-3} 2}(pR)}{\hat{h}^{-}_{l+\frac{D-3} 2}(pR)-\hat{h}^{+}_{l+\frac{D-3} 2}(pR)}\right]\diff p\;,
\end{equation}
where $\gamma$ is a contour that encloses in the counterclockwise direction all the solutions of (\ref{11}) and $d(l)$ is the degeneracy of each eigenvalue, namely
\begin{equation}
d(l)=(2l+D-2)\frac{(l+D-3)!}{l!(D-2)!}\;.
\end{equation}
By deforming the contour of integration $\gamma$ to the imaginary axis and by utilizing the properties \cite{newton,taylor}
\begin{equation}
f_{l}(-k^{\ast})=f^{\ast}_{l}(k)\;,\quad \textrm{and}\quad \hat{h}^{\pm}_{l}(-z)=(-1)^{l}\hat{h}^{\mp}_{l}(z)\;,
\end{equation}
one obtains the following integral representation
\begin{equation}\label{13}
\zeta(s)=\sum_{l=0}^{\infty}d(l)\frac{\sin(\pi s)}{\pi}\int_{m}^{\infty}\left(k^{2}-m^{2}\right)^{-s}\frac{\partial}{\partial k}\log\left[\frac{f_{l}(ik)\hat{h}^{-}_{l+\frac{D-3} 2}(ikR)-f^{\ast}_{l}(ik)\hat{h}^{+}_{l+\frac{D-3} 2}(ikR)}{\hat{h}^{-}_{l+\frac{D-3} 2}(ikR)-\hat{h}^{+}_{l+\frac{D-3} 2}(ikR)}\right]\diff k .
\end{equation}
Due to the asymptotic behavior
\begin{equation}\label{14}
\hat{h}^{\pm}_{l}(ikR)=e^{\mp kR}\left[1+O\left(\frac{1}{kR}\right)\right]\;,
\end{equation}
it is not very difficult to conclude that as $R\to\infty$ one has
\begin{equation}\label{15}
\frac{f_{l}(ik)\hat{h}^{-}_{l+\frac{D-3} 2}(ikR)-f^{\ast}_{l}(ik)\hat{h}^{+}_{l+\frac{D-3} 2}(ikR)}{\hat{h}^{-}_{l+\frac{D-3} 2}(ikR)-\hat{h}^{+}_{l+\frac{D-3} 2}(ikR)}\sim f_{l}(ik)\;.
\end{equation}
According to the remark above, the spectral zeta function associated with the operator ${\cal L}$ acting on scalar functions defined on $\mathbb{R}^{D}$
has the integral representation
\begin{equation}\label{16}
\zeta(s)=\sum_{l=0}^{\infty}d(l)\frac{\sin(\pi s)}{\pi}\int_{m}^{\infty}\left(k^{2}-m^{2}\right)^{-s}\frac{\partial}{\partial k}\log f_{l}(ik)\,\diff k.
\end{equation}
At this point, instead of using the asymptotic expansion of the Jost function to perform the analytic continuation
of $\zeta(s)$ to the region $\Re(s)\leq D/2$ (the approach that was followed in e.g. \cite{bordag96}), we first consider the sum over the angular momentum and then
we focus on the analytic continuation of $\zeta(s)$ \cite{bea15}. By utilizing the Weierstrass factorization theorem \cite{boas54}
we can rewrite the Jost function in terms of the phase shifts $\delta_{l}(q)$ as follows \cite{taylor}
\begin{equation}\label{17}
f_{l}(ik)=\prod_{n}\left(1-\frac{\kappa_{l,n}^{2}}{k^{2}}\right)\exp\left\{-\frac{2}{\pi}\int_{0}^{\infty}\frac{q}{q^{2}+k^{2}}\delta_{l}(q)\diff q\right\}\;,
\end{equation}
where $-\kappa^{2}_{l,n}$ represent the energies of the bound states. It is straightforward to show that once the expression (\ref{17}) is substituted in (\ref{16})
one obtains
\begin{equation}\label{18}
\zeta(s)=\sum_{l=0}^{\infty}d(l)\left\{-\sum_{n}\left[m^{-2s}-\left(m^{2}-\kappa_{l,n}^{2}\right)^{-s}\right]
+\frac{2s}{\pi}\int_{0}^{\infty}\frac{q}{(q^{2}+m^{2})^{s+1}}\delta_{l}(q)\diff q\right\}\;.
\end{equation}
Since the functions $\delta_{l}(q)$ can be expressed in terms of the arctangent \cite{taylor}, they are uniformly bounded in $l$ and $q$.
The last remark, in conjunction with the fact that the function $q(q^{2}+m^{2})^{-s-1}$ is integrable over $\mathbb{R}^{+}$ for $\Re(s)>D/2$, allows us to use
Fubini's theorem to interchange the summation over $l$ and the integration to obtain
\begin{equation}\label{19}
\zeta(s)=\frac{2s}{\pi}\int_{0}^{\infty}\frac{q}{(q^{2}+m^{2})^{s+1}}\delta(q)\diff q\;,
\end{equation}
where, for simplicity, we have assumed that there are no bound states and we have introduced the function
\begin{equation}\label{20}
\delta(q)=\sum_{l=0}^{\infty}d(l) \delta_{l}(q)\;,
\end{equation}
which is well defined for potentials $V(r)$ that vanish for $r\to\infty$ like $r^{-n}$ with $n>D$ \cite{landau}.

\section{Analytic continuation of $\zeta(s)$}

To perform the analytic continuation of the spectral zeta function $\zeta(s)$ in (\ref{19}) to the left of the
line $\Re(s)=D/2$ we need to subtract and then add the large-$q$ asymptotic expansion of the function $\delta(q)$ defined in (\ref{20}).
The desired asymptotic expansion can be found by utilizing the inverse Mellin transform. In fact, by denoting with $\zeta_{0}(s)$ the zeta function
associated with the massless case $m=0$, one can write, from the integral representation (\ref{19}),
\begin{equation}\label{21}
-\frac{\pi}{s}\zeta_{0}\left(-\frac{s}{2}\right)=\int_{0}^{\infty}q^{s-1}\delta(q)\,\diff q\;,
\end{equation}
which is an expression valid for $\Re(s)<-D$ and shows that the function $\delta(q)$ is the Mellin transform of the function on the left-hand-side.
By applying now the inverse transform we obtain
\begin{equation}\label{22}
\delta(q)=-\frac{1}{2i}\int_{c-i\infty}^{c+i\infty}\frac{q^{-s}}{s}\zeta_{0}\left(-\frac{s}{2}\right)\diff s\; ,
\end{equation}
where $c<-D$. The large-$q$ asymptotic expansion of $\delta(q)$ is then simply obtained by shifting the contour of integration to the right.
The poles of the integrand in (\ref{22}) consist of the set of poles of the zeta function $\zeta_{0}\left(-s/2\right)$ and the point $s=0$.
According to the general theory (see e.g. \cite{kirs02b}), the spectral zeta function $\zeta(s)$ has only simple poles at the points
\begin{displaymath}
u=\frac{D-k}{2}\;,\quad k=\{0,1,\ldots,D-1\}\;,\quad\textrm{and}\quad u=-\frac{2l+1}{2}\;,\quad l\in\mathbb{N}_{0}\;,
\end{displaymath}
with residue
\begin{equation}\label{23}
\textrm{Res}\,\zeta(u)=\frac{a_{\frac{D}{2}-u}}{\Gamma(u)}\;,
\end{equation}
where $a_{n/2}$ are the coefficients of the small-$t$ asymptotic expansion of the trace of the heat kernel of the operator from which the zeta function $\zeta(s)$ has
been constructed.

By shifting the contour in (\ref{22}) to the right and by taking into account the poles of the integrand we obtain the
asymptotic expansion
\begin{equation}\label{24}
\delta(q)\sim \pi\sum_{k=0}^{D}\frac{q^{D-k}}{\Gamma\left(\frac{D-k}{2}+1\right)}a_{\frac{k}{2}}+\pi \sum_{l=0}^{\infty}\frac{q^{-2l-1}}{\Gamma\left(-l+\frac{1}{2}\right)}a_{\frac{D+2l+1}{2}}\;.
\end{equation}
Let us mention that the coefficients $a_{n/2}$ of the asymptotic expansion of the trace of the heat kernel for the operator ${\cal P}$ (${\cal L}$ without the mass parameter)
are known for a wide class of potentials $V(r)$. When the potential is a smooth function over its domain, the local coefficients of the asymptotic expansion of the trace of the heat kernel can be computed by utilizing a variety of methods. One of the most straightforward ones
is based on the covariant Fourier transform \cite{avramidi91,nepomechie85}. This method is quite general and allows for the evaluation of the asymptotic expansion of the
local heat kernel for the operator $-\Delta+V({\bf x})$ defined on a smooth compact Riemannian manifold $\mathscr{M}$.
In Euclidean space, by using the integral representation of the Dirac $\delta$-function (see e.g. \cite{toms07}), the heat kernel can be written as
\begin{equation}\label{24ab}
U(t|{\bf x}, {\bf x}')=\int_{\mathbb{R}^{D}}\frac{\diff^{D}k}{(2\pi)^{D}}e^{t\Delta-tV({\bf x})}\cdot e^{ik^{j}(x_{j}-x_{j}')}\;.
\end{equation}
By computing the action of the heat semigroup on the exponential function we obtain
\begin{equation}\label{24ac}
U(t|{\bf x}, {\bf x}')=\int_{\mathbb{R}^{D}}\frac{\diff^{D}k}{(2\pi)^{D}}e^{ik^{j}(x_{j}-x_{j}')}\exp\left\{te^{-ik^{j}(x_{j}-x_{j}')}(\Delta-V({\bf x}))e^{ik^{j}(x_{j}-x_{j}')}\right\}\cdot \mathbb{I}\;,
\end{equation}
with $\mathbb{I}$ denoting the identity function. The operator in the exponential can be rewritten as follows
\begin{equation}\label{24ad}
\exp\left\{te^{-ik^{j}(x_{j}-x_{j}')}(\Delta-V({\bf x}))e^{ik^{j}(x_{j}-x_{j}')}\right\}\cdot \mathbb{I}
=\exp\left\{t\left(-|k|^{2}+2ik^{j}\nabla_{j}+\Delta-V({\bf x})\right)\right\}\cdot \mathbb{I}\;.
\end{equation}
By using (\ref{24ad}) in (\ref{24ac}), by performing the change of variables $k\to k/{\sqrt{t}}$, and by taking the coincidence limit ${\bf x}'\to {\bf x}$
the trace of the heat kernel can be represented as
\begin{equation}\label{24af}
U(t|{\bf x})=\frac{1}{(4\pi t)^{D/2}}\int_{\mathbb{R}^{D}}\frac{\diff^{D}k}{(\pi)^{D/2}}e^{- |k|^{2}}\exp\left\{2i\sqrt{t}k^{j}\nabla_{j}+t\Delta-tV({\bf x})\right\}\cdot \mathbb{I}\;.
\end{equation}
To obtain, from (\ref{24af}), the small-$t$ asymptotic expansion of the trace of the heat kernel we use Trotter's product formula \cite{taylorm11}
to get
\begin{eqnarray}\label{24b}
\lefteqn{\exp\left\{\left(2i\sqrt{t}k^{j}\nabla_{j}+t\Delta-tV({\bf x})\right)\right\}\cdot \mathbb{I}=}\nonumber\\
&&\sum_{m=0}^{\infty}\frac{(-1)^{m}}{m!}t^{m}\left(V({\bf x})\right)^{m}\Bigg\{1+\sum_{n=1}^{\infty}\int_{0}^{1}\diff\tau_{n}\int_{0}^{\tau_{n}}\diff\tau_{n-1}\cdots\int_{0}^{\tau_{2}}\diff\tau_{1}
e^{t\tau_{n}V({\bf x})}\left(2i\sqrt{t}k^{j}\nabla_{j}+t\Delta\right)e^{-t(\tau_{n}-\tau_{n-1})V({\bf x})}\cdots\nonumber \\
&&e^{-t(\tau_{2}-\tau_{1})V({\bf x})}\left(2i\sqrt{t}k^{j}\nabla_{j}+t\Delta\right)e^{-t\tau_{1}V({\bf x})}\Bigg\}\;.
\end{eqnarray}
Once the action of the operators in (\ref{24b}) on the exponential has been computed, one obtains simple multidimensional Gaussian integrals
which can be evaluated by using the formula, for $m\in\mathbb{N}_{0}$,
\begin{equation}
\frac{1}{\pi^{D/2}}\int_{\mathbb{R}^{D}}e^{-|k|^{2}}k_{j_{1}}k_{j_{2}}\cdots k_{j_{n}}\diff^{D}k=
\begin{cases}
0 & \text{if $n=2m+1$}\\
\frac{(2m)!}{2^{2m}m!}\delta_{(j_{1}j_{2}}\cdots\delta_{j_{2m-1}j_{2m})} & \text{if $n=2m$}\;,
\end{cases}
\end{equation}
where $\delta_{ij}$ is the Kronecker delta and the round parentheses denote complete symmetrization of the indexes.

The procedure described above
gives explicit expressions for the coefficients $a_{k}({\bf x})$. For the case of a spherically symmetric potential in $\mathbb{R}^{D}$ considered in this work,
the local coefficients of the asymptotic expansion of the trace of the heat kernel are obtained from $a_{k}(r)$ by subtracting the contribution to the asymptotic expansion of the trace of the heat kernel coming from the Laplacian on $\mathbb{R}^{D}$ without the potential, that is $K_{0}(t|r)=(4\pi t)^{-\frac{D}{2}}$.
In more detail one obtains, for the first few coefficients, \cite{cognola06,parker85}
\begin{eqnarray}\label{24c}
a_{0}(r)&=&0\;,\quad a_{1}(r)=-V(r)\;,\nonumber\\
a_{2}(r)&=&\frac{1}{2}V^{2}(r)-\frac{1}{6}V''(r)-\frac{D-1}{6r}V'(r)\;,\nonumber \\
a_{3}(r)&=&-\frac{1}{60}V^{(4)}(r)-\frac{D-1}{30r}V^{(3)}(r)-\frac{(D-1)(D-3)}{60r^{2}}V''(r)+\frac{(D-1)(D-3)}{60r^{3}}V'(r)+\frac{1}{12}\left(V'(r)\right)^{2}\nonumber\\
&-&\frac{1}{2}V^{3}(r)+\frac{1}{6}V(r)V''(r)+\frac{D-1}{6r}V(r)V'(r)\;.
\end{eqnarray}
Higher order coefficients can be computed with the help of an algebraic computer program. Once the local coefficients $a_{n}(r)$ are found, the
coefficients $a_{n}$ that appear in (\ref{24}) are obtained by integrating the local ones over $\mathbb{R}^{D}$, namely
\begin{equation}
a_{n}=\frac{1}{2^{D-1}\Gamma\left(\frac{D}{2}\right)}\int_{0}^{\infty}r^{D-1}a_{n}(r)\diff r\;.
\end{equation}
Let us mention that the coefficients $a_{k/2}$ of the asymptotic expansion of the trace of the heat kernel for the case of
some non-smooth potential can be found in \cite{gilkey01}.

By subtracting and adding in the integrand of (\ref{19}) $N$ leading terms of the asymptotic expansion (\ref{24}) one obtains
\begin{eqnarray}\label{25}
\zeta(s)=Z_{D}(s)
+2s\sum_{k=0}^{D}\frac{a_{\frac{k}{2}}}{\Gamma\left(\frac{D-k}{2}+1\right)}\int_{0}^{\infty}\frac{q^{1+D-k}}{(q^{2}+m^{2})^{s+1}}\diff q
+2s\sum_{l=0}^{N}\frac{a_{\frac{D+2l+1}{2}}}{\Gamma\left(-l+\frac{1}{2}\right)}\int_{0}^{\infty}\frac{q^{-2l}}{(q^{2}+m^{2})^{s+1}}\diff q\;,
\end{eqnarray}
where, for typographical convenience, we have introduced the function
\begin{equation}\label{25a}
Z_{D}(s)=\frac{2s}{\pi}\int_{0}^{\infty}\frac{q}{(q^{2}+m^{2})^{s+1}}\Bigg\{\delta(q)-\pi\sum_{k=0}^{D}\frac{q^{D-k}}{\Gamma\left(\frac{D-k}{2}+1\right)}a_{\frac{k}{2}}-\pi \sum_{l=0}^{N}\frac{q^{-2l-1}}{\Gamma\left(-l+\frac{1}{2}\right)}a_{\frac{D+2l+1}{2}}\Bigg\}\diff q\;,
\end{equation}
which represents an analytic function of $s$ in the semiplane $\Re(s)>-N-\frac 3 2$. The remaining two integrals in (\ref{25}) can be computed exactly
to lead to the following simple analytically continued expression of the spectral zeta function
\begin{equation}\label{26}
\zeta(s)=Z_{D}(s)+\frac{m^{-2s+D}}{\Gamma(s)}\sum_{k=0}^{D}\Gamma\left(s+\frac{k-D}{2}\right)m^{-k}a_{\frac{k}{2}}
+\frac{m^{-2s-1}}{\Gamma(s)}\sum_{l=0}^{N}\Gamma\left(s+\frac{2l+1}{2}\right)m^{-2l}a_{\frac{D+2l+1}{2}}\;.
\end{equation}
Let us point out that the meromorphic structure of the spectral zeta function is rendered completely manifest in the last formula
and all the information about its simple poles is encoded in the terms of the last two sums.

\section{Functional determinant and Casimir energy}

Once the analytic continuation of the spectral zeta function $\zeta(s)$ is known we can compute, in particular, the zeta regularized functional
determinant of the operator $\mathcal{L}$ according to the formula (\ref{4}), and the Casimir energy of a scalar field under the influence of
the potential $V(r)$ by using the formula given in (\ref{3}).

For the purpose of evaluating the functional determinant, we set $N=-1$ in (\ref{26}), which means the second sum is not needed. We obtain
\begin{equation}\label{27}
\zeta(s)=Z_{D}(s)+\frac{m^{-2s+D}}{\Gamma(s)}\sum_{k=0}^{D}\Gamma\left(s+\frac{k-D}{2}\right)m^{-k}a_{\frac{k}{2}}
\;.
\end{equation}
It this way the resulting expression is valid in the semiplane $\Re(s)>-\frac 1 2$ containing the
point $s=0$ and it will allow us to obtain an expression for $\zeta'(0)$. Since $Z_{D}(s)$ in (\ref{25a}) is an analytic function of
$s$ about $s=0$, we can simply substitute $s=0$ in $Z_{D}'(s)$ to obtain
\begin{equation}\label{27a}
Z_{D}'(0)=\frac{2}{\pi}\int_{0}^{\infty}\frac{q}{q^{2}+m^{2}}\left\{\delta(q)-\pi\sum_{k=0}^{D}\frac{q^{D-k}}{\Gamma\left(\frac{D-k}{2}+1\right)}a_{\frac{k}{2}}\right\}dq .
\end{equation}
By differentiating the sum in (\ref{27}) we find
\begin{eqnarray}\label{28}
\lefteqn{\frac{\partial}{\partial s}\left[\frac{m^{-2s+D}}{\Gamma(s)}\sum_{k=0}^{D}\Gamma\left(s+\frac{k-D}{2}\right)m^{-k}a_{\frac{k}{2}}\right]=}\nonumber\\
&&\frac{m^{-2s+D}}{\Gamma(s)}\sum_{k=0}^{D}\Gamma\left(s+\frac{k-D}{2}\right)m^{-k}a_{\frac{k}{2}}\left[-\log m^{2}-\Psi(s)+\Psi\left(s+\frac{k-D}{2}\right)\right]\; ,
\end{eqnarray}
with the psi-function $\Psi (s) = \frac d {ds} \ln \Gamma (s).$
In order to explicitly evaluate the expression in (\ref{28}) at the point $s=0$ it is convenient to distinguish between even and odd values of the dimension $D$.
When $D=2M$, the derivative in (\ref{28}) can be written, by separating the contributions of the even values of $k$ from the ones of the odd values of $k$, as
\begin{eqnarray}\label{29}
\lefteqn{\frac{\partial}{\partial s}\left[\frac{m^{-2s+D}}{\Gamma(s)}\sum_{k=0}^{D}\Gamma\left(s+\frac{k-D}{2}\right)m^{-k}a_{\frac{k}{2}}\right]=}\nonumber\\
&&\frac{m^{-2s+2M}}{\Gamma(s)}\sum_{j=0}^{M}\Gamma\left(s-M+j\right)m^{-2j}a_{j}\left[-\log m^{2}-\Psi(s)+\Psi\left(s-M+j\right)\right]+\nonumber\\
&&\frac{m^{-2s+2M}}{\Gamma(s)}\sum_{j=0}^{M-1}\Gamma\left(s-M+j+\frac{1}{2}\right)m^{-2j-1}a_{j+\frac{1}{2}}\left[-\log m^{2}-\Psi(s)+\Psi\left(s-M+j+\frac{1}{2}\right)\right]\;.
\end{eqnarray}
The formula obtained in (\ref{29}) is suitable for evaluating the derivative at $s=0$ of the sum in (\ref{27}). In fact, by noticing that
\begin{equation}\label{30}
-\Psi(s)+\Psi\left(s-M+j\right)=\sum_{k=1}^{M-j}\frac{1}{k-s}\;, \quad\textrm{and}\quad \frac{\Psi(s)}{\Gamma(s)}=-1+O(s^{2})\;,
\end{equation}
we have that
\begin{eqnarray}\label{31}
\lefteqn{\frac{\partial}{\partial s}\left[\frac{m^{-2s+D}}{\Gamma(s)}\sum_{k=0}^{D}\Gamma\left(s+\frac{k-D}{2}\right)m^{-k}a_{\frac{k}{2}}\right]\Bigg|_{s=0}=}\nonumber\\
&&m^{2M}\sum_{j=0}^{M}\frac{(-1)^{M-j}}{(M-j)!}m^{-2j}a_{j}\left(-\log m^{2}+H_{M-j}\right)+m^{2M}\sum_{j=0}^{M-1}\Gamma\left(-M+j+\frac{1}{2}\right)m^{-2j-1}a_{j+\frac{1}{2}}\;,
\end{eqnarray}
where we have denoted by $H_{n}$ the $n$-th harmonic number.
When $D=2M+1$, instead, the derivative in (\ref{28}) can be expressed, by separating once again the contributions of the even values of $k$ from the ones of the odd values of $k$, as
\begin{eqnarray}\label{32}
\lefteqn{\frac{\partial}{\partial s}\left[\frac{m^{-2s+D}}{\Gamma(s)}\sum_{k=0}^{D}\Gamma\left(s+\frac{k-D}{2}\right)m^{-k}a_{\frac{k}{2}}\right]=}\nonumber\\
&&\frac{m^{-2s+2M+1}}{\Gamma(s)}\sum_{j=0}^{M}\Gamma\left(s-M+j-\frac{1}{2}\right)m^{-2j}a_{j}\left[-\log m^{2}-\Psi(s)+\Psi\left(s-M+j-\frac{1}{2}\right)\right]+\nonumber\\
&&\frac{m^{-2s+2M+1}}{\Gamma(s)}\sum_{j=0}^{M}\Gamma\left(s-M+j\right)m^{-2j-1}a_{j+\frac{1}{2}}\left[-\log m^{2}-\Psi(s)+\Psi\left(s-M+j\right)\right]\;.
\end{eqnarray}
By utilizing the results in (\ref{30}) we obtain, for odd values of $D$, the following expression
\begin{eqnarray}\label{33}
\lefteqn{\frac{\partial}{\partial s}\left[\frac{m^{-2s+D}}{\Gamma(s)}\sum_{k=0}^{D}\Gamma\left(s+\frac{k-D}{2}\right)m^{-k}a_{\frac{k}{2}}\right]\Bigg|_{s=0}=}\nonumber\\
&&m^{2M+1}\sum_{j=0}^{M}\frac{(-1)^{M-j}}{(M-j)!}m^{-2j-1}a_{j+\frac{1}{2}}\left(-\log m^{2}+H_{M-j}\right)+m^{2M+1}\sum_{j=0}^{M}\Gamma\left(-M+j-\frac{1}{2}\right)m^{-2j}a_{j}\;.
\end{eqnarray}

By combining the results obtained in (\ref{27a}), (\ref{31}), and (\ref{33}) we have the following expressions for the derivative
of the spectral zeta function at $s=0$
\begin{eqnarray}\label{35}
\lefteqn{\zeta'(0)=\frac{2}{\pi}\int_{0}^{\infty}\frac{q}{q^{2}+m^{2}}\left\{\delta(q)-\pi\sum_{k=0}^{2M}\frac{q^{2M-k}}{\Gamma\left(\frac{2M-k}{2}+1\right)}a_{\frac{k}{2}}\right\}dq}\nonumber\\
&&+m^{2M}\sum_{j=0}^{M}\frac{(-1)^{M-j}}{(M-j)!}m^{-2j}a_{j}\left(-\log m^{2}+H_{M-j}\right)+m^{2M}\sum_{j=0}^{M-1}\Gamma\left(-M+j+\frac{1}{2}\right)m^{-2j-1}a_{j+\frac{1}{2}}\; ,
\end{eqnarray}
valid when $D$ is even, and
\begin{eqnarray}\label{36}
\lefteqn{\zeta'(0)=\frac{2}{\pi}\int_{0}^{\infty}\frac{q}{q^{2}+m^{2}}\left\{\delta(q)-\pi\sum_{k=0}^{2M+1}\frac{q^{2M+1-k}}{\Gamma\left(\frac{2M+1-k}{2}+1\right)}a_{\frac{k}{2}}\right\}dq}\nonumber\\
&&+m^{2M}\sum_{j=0}^{M}\frac{(-1)^{M-j}}{(M-j)!}m^{-2j}a_{j+\frac{1}{2}}\left(-\log m^{2}+H_{M-j}\right)+m^{2M+1}\sum_{j=0}^{M}\Gamma\left(-M+j-\frac{1}{2}\right)m^{-2j}a_{j}\; ,
\end{eqnarray}
valid, instead, when $D$ is odd. The functional determinant of $\mathcal{L}$ is then found from (\ref{35}) or (\ref{36}) by using (\ref{4}).

The Casimir energy of the scalar field can be computed by utilizing the expression displayed in (\ref{3}). To this end we set $s=\epsilon-1/2$
in the spectral zeta function (\ref{27}) and compute the expansion of the resulting expression as $\epsilon\to 0$.
By setting $N=0$ in (\ref{25a}) it is not very difficult to realize that $Z_{D}(s)$ becomes an analytic function in the semiplane $\Re(s)>-\frac 3 2$ and
as $\epsilon\to 0$ we get
\begin{equation}\label{37}
Z_{D}\left(-\frac{1}{2}\right)=-\frac{1}{\pi}\int_{0}^{\infty}\frac{q}{\sqrt{q^{2}+m^{2}}}\Bigg\{\delta(q)-\pi\sum_{k=0}^{D+1}\frac{q^{D-k}}{\Gamma\left(\frac{D-k}{2}+1\right)}a_{\frac{k}{2}}\Bigg\}\diff q\;.
\end{equation}

In order to obtain the small-$\epsilon$ expansion of the sum that appears in (\ref{27}), it is, once again, convenient to distinguish between even
and odd values of the dimension $D$. In particular, for $D=2M$ we separate the contributions coming from even and odd values of $k$ to obtain
\begin{eqnarray}\label{38}
\lefteqn{\frac{m^{-2\epsilon+1+2M}}{\Gamma\left(\epsilon-\frac{1}{2}\right)}\sum_{k=0}^{2M+1}\Gamma\left(\epsilon+\frac{k-2M-1}{2}\right)m^{-k}a_{\frac{k}{2}}=
-\frac{m^{2M+1}}{2\sqrt{\pi}}\sum_{j=0}^{M}\Gamma\left(j-M-\frac{1}{2}\right)m^{-2j}a_{j}}\nonumber\\
&&-\frac{m^{2M+1}}{2\sqrt{\pi}}\sum_{j=0}^{M}\frac{(-1)^{M-j}}{(M-j)!}m^{-2j-1}a_{j+\frac{1}{2}}\left[\frac{1}{\epsilon}+\Psi(1+M-j)+\log m^{2}
+\Psi\left(-\frac{1}{2}\right)\right]+O(\epsilon)\;.
\end{eqnarray}
By using the relations
\begin{equation}\label{39}
\Psi(1+M-j)=-\gamma+H_{M-j}\;,\quad\textrm{and}\quad \Psi\left(-\frac{1}{2}\right)=2-2\log 2-\gamma\;,
\end{equation}
one can rewrite (\ref{38}) as
\begin{eqnarray}\label{40}
\lefteqn{\frac{m^{-2\epsilon+1+2M}}{\Gamma\left(\epsilon-\frac{1}{2}\right)}\sum_{k=0}^{2M+1}\Gamma\left(\epsilon+\frac{k-2M-1}{2}\right)m^{-k}a_{\frac{k}{2}}=
-\frac{m^{2M+1}}{2\sqrt{\pi}\epsilon}\sum_{j=0}^{M}\frac{(-1)^{M-j}}{(M-j)!}m^{-2j-1}a_{j+\frac{1}{2}}}\nonumber\\
&&-\frac{m^{2M+1}}{2\sqrt{\pi}}\left\{\sum_{j=0}^{M}\Gamma\left(j-M-\frac{1}{2}\right)m^{-2j}a_{j}+
\sum_{j=0}^{M}\frac{(-1)^{M-j}}{(M-j)!}m^{-2j-1}a_{j+\frac{1}{2}}\left[H_{M-j}-2-\log\left(\frac{m^{2}}{4}\right)\right]\right\}+O(\epsilon)\;.\;\;\;\;\;\;\;\;
\end{eqnarray}
When $D=2M+1$, the small-$\epsilon$ expansion of the sum in (\ref{27}) can be expressed, separating
again the contributions of even values of $k$ from the contributions of odd values of $k$, as
\begin{eqnarray}\label{41}
\lefteqn{\frac{m^{-2\epsilon+2+2M}}{\Gamma\left(\epsilon-\frac{1}{2}\right)}\sum_{k=0}^{2M+2}\Gamma\left(\epsilon+\frac{k-2M-2}{2}\right)m^{-k}a_{\frac{k}{2}}=
-\frac{m^{2M+2}}{2\sqrt{\pi}}\sum_{j=0}^{M}\Gamma\left(j-M-\frac{1}{2}\right)m^{-2j-1}a_{j+\frac{1}{2}}}\nonumber\\
&&-\frac{m^{2M+2}}{2\sqrt{\pi}}\sum_{j=0}^{M+1}\frac{(-1)^{M+1-j}}{(M+1-j)!}m^{-2j}a_{j}\left[\frac{1}{\epsilon}+\Psi(2+M-j)+\log m^{2}
+\Psi\left(-\frac{1}{2}\right)\right]+O(\epsilon)\;.
\end{eqnarray}
By using the results in (\ref{39}) we have
\begin{eqnarray}\label{42}
\lefteqn{\frac{m^{-2\epsilon+2+2M}}{\Gamma\left(\epsilon-\frac{1}{2}\right)}\sum_{k=0}^{2M+2}\Gamma\left(\epsilon+\frac{k-2M-2}{2}\right)m^{-k}a_{\frac{k}{2}}=
-\frac{m^{2M+2}}{2\sqrt{\pi}\epsilon}\sum_{j=0}^{M+1}\frac{(-1)^{M+1-j}}{(M+1-j)!}m^{-2j}a_{j}}\nonumber\\
&&-\frac{m^{2M+2}}{2\sqrt{\pi}}\left\{\sum_{j=0}^{M}\Gamma\left(j-M-\frac{1}{2}\right)m^{-2j-1}a_{j+\frac{1}{2}}+
\sum_{j=0}^{M+1}\frac{(-1)^{M+1-j}}{(M+1-j)!}m^{-2j}a_{j}\left[H_{M+1-j}-2-\log\left(\frac{m^{2}}{4}\right)\right]\right\}+O(\epsilon)\;.\nonumber\\
\end{eqnarray}

The formula (\ref{3}) and the results (\ref{37}), (\ref{40}), and (\ref{42}) provide the following expressions for the Casimir energy
of a scalar field propagating in $\mathbb{R}^{D}$ under the presence of a spherically symmetric potential
\begin{eqnarray}\label{43}
\lefteqn{E_{\textrm{Cas}}(\epsilon)=-\frac{m^{2M}}{4\sqrt{\pi}}\frac{1}{\epsilon}\sum_{j=0}^{M}\frac{(-1)^{M-j}}{(M-j)!}m^{-2j}a_{j+\frac{1}{2}} +\frac{1}{2}Z_{2M}\left(-\frac{1}{2}\right)}\nonumber\\
&&-\frac{m^{2M+1}}{4\sqrt{\pi}}\left[\sum_{j=0}^{M}\Gamma\left(j-M-\frac{1}{2}\right)m^{-2j}a_{j}-\sum_{j=0}^{M}\frac{(-1)^{M-j}}{(M-j)!}m^{-2j-1}a_{j+\frac{1}{2}}\left(2+\ln\left(\frac{m^{2}}{4\mu^{2}}\right)-H_{M-j}\right)\right]+O(\epsilon)\;,\nonumber\\
\end{eqnarray}
which is valid when $D=2M$, and
\begin{eqnarray}\label{44}
\lefteqn{E_{\textrm{Cas}}(\epsilon)=-\frac{m^{2M+2}}{4\sqrt{\pi}}\frac{1}{\epsilon}\sum_{j=0}^{M+1}\frac{(-1)^{M+1-j}}{(M+1-j)!}m^{-2j}a_{j} +\frac{1}{2}Z_{2M+1}\left(-\frac{1}{2}\right)}\nonumber\\
&&-\frac{m^{2M+2}}{4\sqrt{\pi}}\left[\sum_{j=0}^{M}\Gamma\left(j-M-\frac{1}{2}\right)m^{-2j-1}a_{j+\frac{1}{2}}-\sum_{j=0}^{M+1}\frac{(-1)^{M+1-j}}{(M+1-j)!}m^{-2j}a_{j}\left(2+\ln\left(\frac{m^{2}}{4\mu^{2}}\right)-H_{M+1-j}\right)\right]+O(\epsilon)\;,\nonumber\\
\end{eqnarray}
valid, instead, for $D=2M+1$.

As it is to be expected, the Casimir energy given above needs to be renormalized. We impose the following renormalization condition
\begin{equation}\label{45}
\lim_{m\to\infty}E_{\textrm{Cas}}^{(\textrm{ren})}=0\;,
\end{equation}
which simply states that the quantum fluctuations must vanish in the classical limit, namely, when the mass of the field $m\to\infty$.
By applying this renormalization condition to the expressions for the Casimir energy in (\ref{43}) and (\ref{44}) we get the very simple result
\begin{equation}\label{46}
E_{\textrm{Cas}}^{(\textrm{ren})}=\frac{1}{2}Z_{D}\left(-\frac{1}{2}\right)\;,
\end{equation}
which is valid for both even and odd values of the dimension $D$. The divergent part of the Casimir energy, denoted by $E_{\textrm{Cas}}^{(\textrm{div})}$,
is defined according to the expression
\begin{eqnarray}\label{47}
E_{\textrm{Cas}}(\epsilon)=E_{\textrm{Cas}}^{(\textrm{ren})}+E_{\textrm{Cas}}^{(\textrm{div})}(\epsilon)\;,
\end{eqnarray}
and reads
\begin{eqnarray}\label{48}
E_{\textrm{Cas}}^{(\textrm{div})}&=&-\frac{m^{2M}}{4\sqrt{\pi}}\sum_{j=0}^{M}\frac{(-1)^{M-j}}{(M-j)!}m^{-2j}a_{j+\frac{1}{2}}
\left[\frac{1}{\epsilon}-2-\ln\left(\frac{m^{2}}{4\mu^{2}}\right)+H_{M-j}\right]\nonumber\\
&-&\frac{m^{2M+1}}{4\sqrt{\pi}}\sum_{j=0}^{M}\Gamma\left(j-M-\frac{1}{2}\right)m^{-2j}a_{j}\;,
\end{eqnarray}
when $D=2M$, and
\begin{eqnarray}\label{49}
E_{\textrm{Cas}}^{(\textrm{div})}&=&-\frac{m^{2M+2}}{4\sqrt{\pi}}\sum_{j=0}^{M+1}\frac{(-1)^{M+1-j}}{(M+1-j)!}m^{-2j}a_{j+\frac{1}{2}}
\left[\frac{1}{\epsilon}-2-\ln\left(\frac{m^{2}}{4\mu^{2}}\right)+H_{M+1-j}\right]\nonumber\\
&-&\frac{m^{2M+1}}{4\sqrt{\pi}}\sum_{j=0}^{M}\Gamma\left(j-M-\frac{1}{2}\right)m^{-2j}a_{j+\frac{1}{2}}\; ,
\end{eqnarray}
when $D=2M+1$.

The results obtained in this section for the functional determinant and the Casimir energy are very general and are valid
for any dimension $D$ and for any radially symmetric potential $V(r)$ that decays sufficiently fast as $r\to\infty$.
Obviously, the knowledge of the phase shift $\delta(q)$ is required in order to obtain explicit results for particular cases.

To conclude, let us provide final results for the determinant and the Casimir energy in $D=2$ and $D=3$ dimensions for the case of a smooth potential.
From (\ref{35}), in $D=2$, we find using known results for the heat kernel coefficients \cite{gilkey95,kirs02b}
\beq
\zeta ' _{D=2} (0)  = \frac 2 \pi \int\limits_0^\infty \frac q {q^2+m^2} \left\{ \delta (q) + \frac \pi 2 \int\limits_0^\infty r V(r) dr \right\} dq + \frac 1 2 \ln m^2 \,\, \int\limits_0^\infty r V(r) dr\;.
\eeq
Similarly, from (\ref{36}), in $D=3$,
\beq
\zeta '_{D=3} (0)  = \frac 2 \pi \int\limits_0^\infty \frac q {q^2+m^2} \left\{ \delta (q) + q\,\, \int\limits_0^\infty r^2 V(r) dr \right\} dq + m\,\,  \int\limits_0^\infty r^2 V(r) dr\; ,
\eeq
and again from (\ref{35}), in $D=4$,
\beq
\zeta '_{D=4} (0) & =& \frac 2 \pi \int\limits_0^\infty \frac q {q^2+m^2} \left\{ \delta (q) + \frac \pi 8 \,\, q^2\,\, \int\limits_0^\infty r^3 V(r) dr -\frac \pi {16} \int\limits_0^\infty
r^3 V^2 (r) dr \right\} dq \nonumber\\
& & +\frac 1 8 m^2\,\,(1-\ln m^2) \,\,  \int\limits_0^\infty r^3 V(r) dr-\frac 1 {16} \ln m^2 \int\limits_0^\infty r^3 V^2 (r) dr\; .
\eeq
The Casimir energies are obtained from (\ref{37}) and read
\beq
E_{Cas,D=2} ^{(ren)} &=& - \frac 1 {2\pi} \int\limits_0^\infty \frac q {\sqrt{q^2+m^2}} \left\{ \delta (q) + \frac \pi 2 \int\limits_0^\infty r V(r) dr\right\} dq\;,\\
E_{Cas,D=3} ^{(ren)} &=& - \frac 1 {2\pi} \int\limits_0^\infty \frac q {\sqrt{q^2+m^2}} \left\{ \delta (q) + q \,\, \int\limits_0^\infty r^2 V(r) dr - \frac 1 {4q} \int\limits_0^\infty r^2 V^2 (r) dr\right\} dq \;,\\
E_{Cas,D=4} ^{(ren)} &=& - \frac 1 {2\pi} \int\limits_0^\infty \frac q {\sqrt{q^2+m^2}} \left\{ \delta (q) +\frac \pi 8  q^2 \,\, \int\limits_0^\infty r^3 V(r) dr - \frac \pi {16} \int\limits_0^\infty r^3 V^2 (r) dr\right\} dq \;.
\eeq
Higher dimensional results are obtained with ease.

\section{Concluding Remarks}

In this paper we have performed the analytic continuation of the spectral zeta function associated with the Laplace operator acting on scalar
functions on $\mathbb{R}^{D}$ endowed with a spherically symmetric potential $V(r)$ which decays sufficiently fast as $r\to\infty$.
We have used the analytically continued expression of the spectral zeta function to compute the functional determinant of the
operator and the Casimir energy associated with the scalar field.
The approach used in this work to study the spectral zeta function is based on the phase shift which is obtained from the Jost function.
The analytic continuation of $\zeta(s)$ was performed by subtracting and adding from a suitable integral representation of the spectral zeta function
a finite number of leading terms of the asymptotic expansion of the phase shift. One of the advantages of this approach is that
the asymptotic expansion of the phase shift is intimately related to the small-$t$ asymptotic expansion of the trace of the heat kernel
associated with the Laplace operator and, hence, is immediately available for a wide range of cases, without the need for additional calculations.
The technically important advantage of the method used in this paper to analyze the spectral zeta function stems from
performing the sum over the angular momenta in (\ref{18}) before computing the integral over the imaginary frequencies $q$.
This leads to the simple integral representation of $\zeta(s)$ given in (\ref{19}), which, in turn, provides very straightforward
results for the functional determinant and the Casimir energy.

By reading the previous section it becomes clear that in order to obtain explicit results one needs to know the phase shift $\delta(q)$.
In general the phase shift is not known explicitly, however there are simple, although important, cases where $\delta(q)$ can be found explicitly.
A few examples of these cases in two and three dimensional settings have been discussed in \cite{bea15}. Obviously, it would be very interesting to
find higher dimensional cases where the phase shift is known explicitly and to apply the method presented here. This would provide specific results for the
functional determinant and the Casimir energy.
It would be also interesting to extend the method presented in this work to the class of singular problems in which the
small-$t$ asymptotic expansion of the trace of the heat kernel contains logarithmic terms \cite{bru}. In this case the spectral zeta function
acquires additional, non-standard, poles which would modify the asymptotic expansion of $\delta(q)$ in (\ref{24}).
Finally, it would be relevant to apply the underlying ideas to higher spin particles \cite{bord99-60-105019,bord03-67-065001,bord03-68-065026}, where even more dramatic simplifications are expected to occur.

\section{Acknowledgments}
KK is very grateful to Michael Bordag for fruitful discussions on the subject.


\begin{thebibliography}{99}

\bibitem{adki83-228-552}
G.S. Adkins, C.R. Nappi, and E.~Witten.
\newblock Static properties of nucleons in the {S}kyrme model.
\newblock {\em Nucl. Phys.}, B228:552--566, 1983.

\bibitem{ambj85-256-434}
J.~Ambjorn and V.A. Rubakov.
\newblock Classical versus semiclassical electroweak decay of a techniskyrmion.
\newblock {\em Nucl. Phys.}, B256:434--448, 1985.

\bibitem{avramidi91} I.G. Avramidi.
\newblock The covariant technique for calculation of one-loop effective action.
\newblock \emph{Nucl. Phys.} B355:712, 1991.

\bibitem{bea15} M. Beauregard, M. Bordag, and K. Kirsten.
\newblock Casimir energies in spherically symmetric background potentials revisited.
\newblock {\em J. Phys. A: Math. Theor.}, 48:095401, 2015.

\bibitem{boas54} R.P. Boas.
\newblock {\em Entire functions}.
\newblock Academic Press Inc., New York, NY, 1954

\bibitem{bord03-67-065001} M. Bordag.
\newblock On the vacuum energy of a color magnetic vortex.
\newblock {\em Phys. Rev.}, D67:065001, 2003.

\bibitem{bord03-68-065026} M. Bordag and I. Drozdov.
\newblock Fermionic vacuum energy from a Nielsen-Olesen vortex.
\newblock {\em Phys. Rev.}, D68:065026, 2003.

\bibitem{bordag96} M. Bordag and K. Kirsten.
\newblock Vacuum energy in a spherically symmetric background field.
\newblock {\em Phys. Rev.}, D53:5753--5760, 1996.

\bibitem{bord99-60-105019}
M.~Bordag and K.~Kirsten.
\newblock The ground state energy of a spinor field in the background of a finite radius flux tube.
\newblock {\em Phys. Rev.}, D60:105019, 1999.

\bibitem{bord09b}
M.~Bordag, G.L. Klimchitskaya, U.~Mohideen, and V.M. Mostepanenko.
\newblock {\em Advances in the Casimir effect}.
\newblock Oxford Science Publications, Oxford, 2009.

\bibitem{bru} J. Br\"{u}ning and R. Seeley.
\newblock The resolvent expansion for second order regular singular operators.
\newblock {\em J. Func. Anal.}, 73:396, 1987.

\bibitem{bytsenko03} A.A. Bytsenko, G. Cognola, E. Elizalde, V. Moretti, and S. Zerbini.
\newblock {\em Analytic Aspects of Quantum Fields}.
\newblock World Scientific Publishing, Singapore, 2003.

\bibitem{cognola06} G. Cognola, E. Elizalde, and S. Zerbini.
\newblock Heat-kernel expansion on noncompact domains and a
generalized zeta-function regularization procedure.
\newblock \emph{J. Math. Phys}, 47:083516, 2006.

\bibitem{dowk76-13-3224}
J.S. Dowker and R. Critchley.
\newblock Effective Lagrangian and energy momentum tensor in de Sitter space.
\newblock {\em Phys. Rev.} D13:3224--3232, 1976.

\bibitem{dunne09} G.V. Dunne and K. Kirsten.
\newblock Simplified vacuum energy expressions for radial backgrounds and domain walls.
\newblock {\em J. Phys. A: Math. Theor.}, 42:075402, 2009.

\bibitem{eila86-56-1331}
G.~Eilam, D.~Klabucar, and A.~Stern.
\newblock Skyrmion solutions to the {W}einberg-{S}alam model.
\newblock {\em Phys. Rev. Lett.}, 56:1331--1334, 1986.

\bibitem{elizalde94} E. Elizalde, S.D. Odintsov, A. Romeo, A. Bytsenko, and S. Zerbini.
\newblock {\em Zeta Regularization Techniques with Applications}.
\newblock World Scientific, Singapore, 1994.

\bibitem{frie77-15-1694}
R.~Friedberg and T.D. Lee.
\newblock Fermion field nontopological solitons. {I}.
\newblock {\em Phys. Rev.}, D15:1694--1711, 1977.

\bibitem{frie77-16-1096}
R.~Friedberg and T.D. Lee.
\newblock Fermion field nontopological solitons. {I}{I}. {M}odels for hadrons.
\newblock {\em Phys. Rev.}, D16:1096--1118, 1977.

\bibitem{gilkey95} P.B. Gilkey.
\newblock {\em Invariance Theory, the Heat Equation, and the Atiyah-Singer Index Theorem}.
\newblock CRC Press, Boca Raton, FL, 1995.

\bibitem{gilkey01} P.B. Gilkey, K. Kirsten, and D.V. Vassilevich.
\newblock Heat trace asymptotics with transmittal boundary conditions and quantum brane world scenario.
\newblock {\em Nucl. Phys.} B601:125--148, 2001.

\bibitem{gips84-231-365}
J.M. Gipson.
\newblock Quasi-solitons in the strongly coupled {H}iggs sector of the standard
  model.
\newblock {\em Nucl. Phys.}, B231:365--385, 1984.

\bibitem{gips81-183-524}
J.M. Gipson and C.-H. Tze.
\newblock Possible heavy solitons in the strongly coupled {H}iggs sector.
\newblock {\em Nucl. Phys.}, B183:524--546, 1981.

\bibitem{hawk77-55-133}
S.W. Hawking.
\newblock Zeta function regularization of path integrals in curved space-time.
\newblock {\em Commun. Math. Phys.}, 55:133-148, 1977.

\bibitem{landau} L.D. Landau and E.M. Lifshitz.
\newblock {\em Quantum mechanics: non-relativistic theory. Course of Theoretical Physics, Vol. 3}.
\newblock Addison-Wesley Publishing Co., Inc., Reading, MA, 1958

\bibitem{mina49} S. Minakshisundaram and A. Pleijel.
\newblock Some properties of the eigenfunctions of the Laplace-operator on Riemannian manifolds.
\newblock \emph{Canad. J. Math.}, 1:242--256, 1949.

\bibitem{nepomechie85} R.I. Nepomechie.
\newblock Calculating heat kernels.
\newblock \emph{Phys. Rev.}, D31:3291, 1985.

\bibitem{newton}
R.G. Newton
\newblock {\em Scattering Theory of Waves and Particles}, Text and Monographs in Physics.
\newblock Springer-Verlag, New York, NY, 1982.

\bibitem{kirs02b}
K.~Kirsten.
\newblock {\em Spectral Functions in Mathematics and Physics}.
\newblock Chapman\&Hall/CRC, Boca Raton, FL, 2002.

\bibitem{kirs03-308-502}
K. Kirsten and A.J. McKane.
\newblock Functional determinants by contour integration methods.
\newblock {\em Ann. Phys.}, 308:502--527, 2003.

\bibitem{klin84-30-2212}
F.R. Klinkhamer and N.S. Manton.
\newblock A saddle point solution in the {W}einberg-{S}alam theory.
\newblock {\em Phys. Rev.}, D30:2212--2220, 1984.

\bibitem{moss02-632-173}
I.G. Moss and W. Naylor.
\newblock Effective action for bubble nucleation rates.
\newblock {\em Nucl. Phys. B}, 632:173--88, 2002.

\bibitem{parker85} L. Parker and D.J. Toms.
\newblock New form for the coincidence limit of the Feynman propagator, or heat kernel,
in curved spacetime.
\newblock \emph{Phys. Rev.}, D31:953, 1985.

\bibitem{poly74-20-194}
A.M. Polyakov.
\newblock Particle spectrum in quantum field theory.
\newblock {\em JETP Lett.}, 20:194--195, 1974.

\bibitem{ray71} D.B. Ray and I.M. Singer.
\newblock R-torsion and the Laplacian on Riemannian manifolds.
\newblock {\em Advances in Math.}, 98:154--177, 1971.

\bibitem{seeley} R.T. Seeley.
\newblock Complex powers of an elliptic operator, Singular Integrals, Chicago 1966.
\newblock {Proc. Sympos. Pure Math.}, 10:288--307, American Mathematical Society, Providence, RI, 1968.

\bibitem{skyr61-260-127}
T.H.R. Skyrme.
\newblock A nonlinear field theory.
\newblock {\em Proc. Roy. Soc. Lond.}, A260:127--138, 1961.

\bibitem{skyr62-31-556}
T.H.R. Skyrme.
\newblock A unified field theory of mesons and baryons.
\newblock {\em Nucl. Phys.}, B31:556--569, 1962.

\bibitem{taylorm11}
M.E. Taylor.
\newblock {\em Partial differential equations II. Qualitative studies of linear equations}
\newblock Applied Mathematical Sciences, {\bf 116}, Springer, New York, (2011)

\bibitem{taylor}
J.R. Taylor.
\newblock {\em Scattering Theory}.
\newblock Wiley, New York, NY, 1972.


\bibitem{toms07} D.J. Toms.
\newblock{\em The Schwinger action principle and effective action}, Cambridge Monographs on Mathematical Physics,
\newblock  Cambridge University Press, Cambridge, UK, 2007

\bibitem{hoof74-79-276}
G.'t~Hooft.
\newblock Magnetic monopoles in unified gauge theories.
\newblock {\em Nucl. Phys.}, B79:276--284, 1974.

\bibitem{voros87} A. Voros.
\newblock Spectral functions, special functions and Selberg zeta function.
\newblock \emph{Commun. Math. Phys.}, 110:439--465, 1987.


\end{thebibliography}
\end{document}